\documentclass[amssymb,prb,twocolumn,showpacs]{revtex4}
\usepackage{amsmath}
\usepackage{graphicx}
\usepackage{verbatim}
\usepackage{graphics}
\usepackage{mathdots}

\makeatletter
\renewcommand*\env@matrix[1][*\c@MaxMatrixCols c]{%
  \hskip -\arraycolsep
  \let\@ifnextchar\new@ifnextchar
  \array{#1}}
\makeatother

\begin{document}

\title{Dynamical polarizability of graphene irradiated by circularly polarized ac electric fields}
\author{Maria Busl$^1$, Gloria Platero$^1$, and Antti-Pekka Jauho$^2$}
\affiliation{$1$ Instituto de Ciencia de Materiales de Madrid, CSIC, Cantoblanco, 28049 Madrid, Spain}
\affiliation{$^2$Center for Nanostructured Graphene (CNG), Department of Micro- and Nanotechnology, DTU Nanotech, Technical University of Denmark, DK-2800 Kongens Lyngby, Denmark}
\date{\today}
\begin{abstract}
\vspace{0.5cm}
We examine the low-energy physics of graphene in the presence of a circularly polarized electric field in the terahertz regime.  Specifically, we derive a general expression  for the dynamical polarizability of graphene irradiated by an ac electric field.  Several approximations are developed that allow one to develop a semianalytical theory for the weak field regime.
The ac field changes qualitatively the single and many electron excitations of graphene: undoped samples may exhibit collective excitations (in contrast to the equilibrium situation), and the properties of the excitations in doped graphene are strongly influenced by the ac field. We also show that the intensity of the external field is the critical control parameter for the stability of these excitations.
\end{abstract}
\maketitle


\section{Introduction}

Graphene is a genuinely two dimensional material whose peculiar properties have received a lot of attention since its first isolation in 2004.\cite{geimSC04, zhangNA05} Structurally, it is a single atom thick layer of graphite, i.e. a two dimensional crystal, that remains stable both when it is deposited over a substrate or when it is suspended. Its electronic properties have attracted huge interest: the low energy excitations are chiral massless Dirac electrons in two dimensions, thereby providing a new platform for testing the basic tenets of solid state physics. This fact, which ultimately arises from the honeycomb structure of the graphene crystal lattice, is responsible for a strikingly different electronic behavior as compared to the conventional two dimensional electron gases (e.g., in semiconductor heterostructures) studied extensively in the laboratory.\cite{guineaRMP09, andoRMP82}

The effect of external fields in the low energy properties of the electric carriers in graphene has been a topic of extensive research since early days, as the discovery or the anomalous Quantum Hall effect witnesses.\cite{geimNA05,zhangNA05} Understanding the behavior of graphene in the presence of electrical and magnetic fields is of major relevance both from a fundamental and an applied point of view. The former, since new exotic behavior may arise in the presence of external fields,
and the latter, because external fields can be used to manipulate its properties, for instance by opening gaps in the electronic spectrum, which is essential for applications in the semiconductor industry.

The effect of radiation on both monolayer and bilayer graphene has been analyzed only recently, and has led to the prediction of a variety of phenomena, such as the photovoltaic Hall effect,\cite{aokiPRB09} metal-insulator transition of graphene,\cite{kibisPRB10} valley-polarized currents in both monolayer and bilayer graphene,\cite{chakrabortyAPL09, chakrabortyNT11} and photoinduced quantum Hall effect in the absence of magnetic fields.\cite{demlerPRB11}
Other theoretical works include the analysis of ac transport properties through graphene ribbons,\cite{fertigPRL11} graphene-based $pn$-junctions,\cite{efetovPRB08} graphene-based Fabry-P\'erot devices\cite{cunibertiPRB10} and the recent proposal of quantum pumping in graphene by an external ac field.\cite{kohlerPRB11}  Experimentally it has been found that a circularly polarized ac field induces a dynamic Hall effect in graphene.\cite{olbrichPRL10}
Several studies have been devoted to the theoretical analysis of the quasienergy spectrum of graphene and graphene dots under ac fields,\cite{calvoAPL11, naumisPRB08, riveraPRB09, zhangNJofPhys09} and the optical properties of graphene have been studied by calculating the optical conductivity.\cite{zhouPRB11}
One of the earliest and yet most important findings in all these studies is that a circularly polarized field induces a band-gap at the Dirac point,
along with dynamical gaps at other momenta, all of which are tunable by the field intensity. This is, however, not the case for a linearly polarized field: there the {\it anisotropic} quasienergy spectrum shows dynamical gaps at non-zero momentum only in certain directions, and especially no gap is induced at the Dirac point.\cite{zhouPRB11,calvoAPL11}

In this paper we study theoretically the effect of a circularly polarized ac electric field in the terahertz regime on the electron excitation spectrum, and on the electron-electron interaction.
The interactions are found to be affected qualitatively by the external field, altering the nature of the single particle excitations as well as the many-particle excitations, both in doped and undoped graphene. Special attention is paid to the existence of a plasmon in {\it undoped} graphene, which is not present in its field-free counterpart. In order to perform this investigation, the natural object to study is the dynamical polarizability, which has already been studied extensively in graphene without an ac field.\cite{shungPRB86, dassarmaPRB07, wunschNJoP06, mishchenkopolPRL08, sabiofsumPRB08, stauberanalyticalpolPRB10, roldandynpolmagPRB09}

The structure of the paper is the following: In Sect.~\ref{sec2}, we briefly introduce the Hamiltonian of graphene in the presence of a circularly polarized electric field, emphasizing the role of Floquet theory in Sect.~\ref{sec2subsec1}, and present several approximations to the single electron Hamiltonian valid for weak fields in Sect.~\ref{sec2subsec2}.
Section~\ref{sec3} is dedicated to the analysis of the dynamical polarizability: we derive a general expression for the polarizability of graphene in an ac electric field in Sect.~\ref{sec3subsec1}, and compare it with the corresponding expression for the two dimensional electron gas.\cite{chinodynscreeningPRB02}
Finally in section \ref{sec3subsec2}, the general formula is combined with the analytical approximations in order to work it out both for the non-interacting system
and for the interacting system in the Random Phase Approximation (RPA).


\section{Single electron properties of graphene under a circularly polarized ac electric field}
\label{sec2}

\subsection{Model and technique}
\label{sec2subsec1}

In the low-energy regime, the Hamiltonian for single electron excitations in graphene is the infamous Dirac Hamiltonian. In order to introduce a time-dependent electric field we choose a gauge in which the latter is represented via a gauge potential $\mathbf{A}(t)$, whose time dependence is that of a single monochromatic and circularly polarized wave of frequency $\Omega$:
\begin{align}
\mathbf{A}(t) &= -\frac{E_0}{\Omega\sqrt{2}}[\hat{x}\sin(\Omega t) - \hat{y}\cos(\Omega t)],
\end{align}
By using a minimal coupling scheme, the Hamiltonian for graphene irradiated by this electric field reads:
\begin{equation}
H(t) = \begin{pmatrix}
0 & k_x - i k_y + iAe^{-i\Omega t}\\
k_x +  i k_y-iA e^{i\Omega t} & 0
\end{pmatrix},
\label{hamilcirc}
\end{equation}
with $A = eE_{0}/(\sqrt{2}\Omega)$ and $v_{\text{F}} = \hbar = 1$.
Here the Hamiltonian is expressed in terms of Bloch states of momentum $\mathbf{k}$, which is defined with respect to one of the valleys.
Note that the electric field does not couple the spin and valley degrees of freedom in graphene, which remain as an extra degeneracy $N_v N_s = 4$. The eigenstates of graphene in the absence of the external field are two-dimensional spinors representing the two components of the unit cell of the honeycomb lattice in graphene, that once diagonalized give rise to two bands (or Dirac cones). In the Dirac Hamiltonian, the pseudospin has a scalar coupling with the momentum, and its eigenstates are those whose pseudospin is either parallel or antiparallel to its momentum. In fact, the mathematical structure of the Hamiltonian coincides with that of an electronic spin coupled through Rashba interaction to a magnetic field. In this analogy,  the momentum in graphene plays the role of the magnetic field, and the pseudospin operator is the analogous to the ordinary spin, both having the same representation in terms of Pauli matrices. This allows one to write the Hamiltonian $H = {\bf \sigma} \cdot {\bf k}$.  In the presence of an external electric field an extra term of the same nature arises in the Hamiltonian, now coupling the pseudospin and the electric field, and inducing transitions between the eigenstates for the isolated system. In a sense the momentum and the electric field are competing dynamically for the direction of the pseudospin, but no compromise can be reached due to the time-dependence of the field, which no longer allows for an analysis of the problem in terms of stationary eigenstates.\\

 To proceed, we apply the Floquet theorem, which is the most suitable way to address time periodic Hamiltonians (detailed accounts can be found in Refs.~[\onlinecite{grifoniPhysRep98, plateroPhysRep04, kohlerPhysRep05}]).
Floquet theory states that for a Hamiltonian that is periodic in time -- $H(t + 2\pi/\Omega) = H(t)$ -- a complete set of solutions of the time-dependent
Schr\"{o}dinger equation
\begin{equation}
H(t)|\psi(t)\rangle = i\frac{d}{dt}|\psi(t)\rangle
\label{sgltime}
\end{equation}
can be written as
\begin{align}
|\psi_{\alpha}(t)\rangle &= e^{-i\epsilon_{\alpha}t}|\phi_{\alpha}(t)\rangle\nonumber\\
|\phi_{\alpha}(t)\rangle &= |\phi_{\alpha}(t+T)\rangle,
\label{sglsol}
\end{align}
where $\alpha$ contains the quantum numbers of the problem and the so-called Floquet index, that we will label as $l$. The role of this index is to classify the different {\it sidebands}, since $\epsilon_{\alpha}$, the quasienergies, are defined $\mod \hbar\Omega$, being related by the simple transformation:
\begin{equation}
\epsilon_{\alpha(l)} = \epsilon_{\alpha(0)} + l \Omega
\label{transformation1}
\end{equation}
In analogy to the Bloch theorem, the quasienergies can be mapped into a first time Brillouin zone, which is $[-\Omega/2, \Omega/2]$, and therefore corresponds to $l = 0$.

The Floquet states $|\phi_{\alpha}(t)\rangle$ have the same periodicity as the driving field (see Eq.\eqref{sglsol})
and can therefore be expanded into a Fourier series:
\begin{equation}
|\phi_{\alpha}(t)\rangle = \sum_{n=-\infty}^{\infty}e^{in\Omega t}|\phi_{\alpha}^n\rangle.
\label{fourierfloquet}
\end{equation}
The Floquet states are also defined in different branches of solutions, being related between them by the transformation:
\begin{equation}				
|\phi_{\alpha(l)}^n\rangle = |\phi_{\alpha(0)}^{l + n}\rangle
\label{transformation2}
\end{equation}
Substituting Eq.\eqref{sglsol} into Eq.\eqref{sgltime} and using Eq.\eqref{fourierfloquet} yields a static eigenvalue equation of the form
\begin{equation}
\sum_m (H^{nm} - n \Omega \delta_{mn}) |\phi_\alpha^m\rangle = \epsilon_\alpha|\phi_\alpha^n\rangle.
\label{floquet}
\end{equation}
Defining now the Floquet Hamiltonian as $H_{\text{F}} ^{nm}= H^{nm} - m\Omega \delta_{mn}$, we see that a significant simplification has been achieved: the time-dependent problem has been transformed to a static problem, and, consequently, one can apply the intuition about equilibrium problems to make statements about a dynamical problem. The resulting equilibrium-like observables derived within this framework have to be understood as time averages over a period of the external field. \\

Let us now apply the Floquet formalism to the Hamiltonian of graphene \eqref{hamilcirc}. In this case, the solutions are characterized by indices $\alpha=(\mathbf{k},\sigma,l)$, being $\sigma = \pm$ the pseudospin index:
\begin{align}
n\Omega\phi_{\alpha}^{n, a} +  (k_x - ik_y)\phi_{\alpha}^{n,b} + i A\phi_{\alpha}^{n+1, b} &= \epsilon_{\alpha}\phi_{\alpha}^{n,a}\nonumber\\
n\Omega\phi_{\alpha}^{n,b} +  (k_x + ik_y)\phi_{\alpha}^{n,a} - i A\phi_{\alpha}^{n-1,a} &= \epsilon_{\alpha}\phi_{\alpha}^{n,b}
\end{align}
Notice that $a$ and $b$ are the indices for the sublattices of the honeycomb lattice. These equations can be written in matrix form, where the infinite Floquet Hamiltonian reads

\begin{widetext}
\begin{equation}
H_{\text{F}} =
\begin{pmatrix}
\ddots & \vdots & \vdots & \vdots & \vdots & \vdots & \vdots & \iddots \\
\cdots & -\Omega &  (k_x - ik_y) & 0 & i A & 0 & 0 & \cdots\\
\cdots &  (k_x + ik_y) & -\Omega & 0 & 0 & 0 & 0 & \cdots\\
\cdots & 0 & 0 & 0 &  (k_x - ik_y) & 0 & i A & \cdots\\
\cdots & -i A & 0 &  (k_x + ik_y) & 0 & 0 & 0 & \cdots\\
\cdots & 0 & 0 & 0 & 0 & \Omega &  (k_x - ik_y) & \cdots\\
\cdots & 0 & 0 & -i A & 0 &  (k_x + ik_y) & \Omega & \cdots\\
\iddots & \vdots & \vdots & \vdots & \vdots & \vdots & \vdots & \ddots
\label{hamilfloquetcirc}
\end{pmatrix}.
\end{equation}
\end{widetext}
The structure of this Hamiltonian deserves a few comments.  The ac field $A$ connects $(2\times2)$ graphene Hamiltonians with energies $n\Omega$ and $(n+1)\Omega$ and so on. Each of these building blocks contributes with its own dispersion relation (that of a Dirac cone) to the energy spectrum, and the field introduces transitions between these cones. These transitions are expressed as anticrossings in the spectrum, and become exact crossings for $A\to0$. It can easily be seen that the anticrossings occur at $|\mathbf{k}|\approx n\Omega/2$, with $n=0,1,2,\dots$. At $|\mathbf{k}|\approx\Omega/2$ e.g., the $(+,0)$ and the $(-,1)$ sideband anticross, which would be a so-called {\it one--photon resonance}.
%

\subsection{Analytical approximations to the single particle Hamiltonian}
\label{sec2subsec2}

\begin{figure}[t!]
\begin{center}
\includegraphics[width=3.4in,clip] {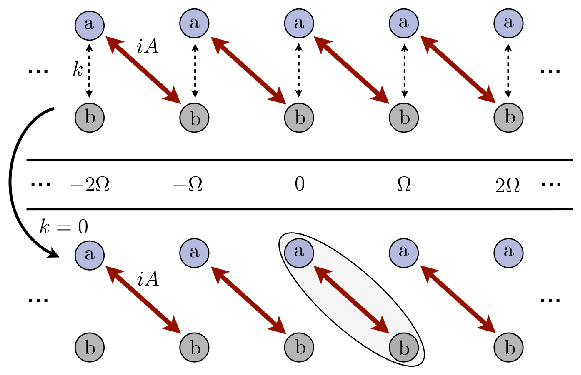}
\end{center}
\caption{\label{chaincirc}\small
Sketch of the Hamiltonian for the circularly polarized field. Note that if $\mathbf{k}=0$, the Hamiltonian breaks up into disconnected two-level systems,
in which site $a_n$ is coupled to site $b_{n+1}$.
}
\end{figure}

The Floquet Hamiltonian, Eq.~(\ref{hamilfloquetcirc}) can be diagonalized numerically in order to analyze the energy spectrum and its features.
However, in order to simplify calculations and to illuminate the main physics, here we resort to analytical approximations which capture the main features whenever the electric field intensity is sufficiently weak.
We will show in Sect.~\ref{sec2subsec3} the full numerical results for the quasienergy spectrum in order to compare it with the analytical approximations.
Before introducing such approximations, it is convenient to project the Hamiltonian \eqref{hamilfloquetcirc} into another basis.
As can be seen in Fig.~\ref{chaincirc}, for $\mathbf{k}=0$, the Floquet chain breaks up into a series of disconnected two-level systems.
We therefore diagonalize the Hamiltonian for the two-level-system, and then write the full Hamiltonian in the resulting basis.
The Hamiltonian for $\mathbf{k}=0$ reads
\begin{multline}
H^{\mathbf{k=0}}_{\text{F}} = \sum_n n\Omega\left[|\phi^{n,a}_{\mathbf{k} = 0}\rangle\langle \phi^{n,a}_{\mathbf{k} = 0}| + |\phi^{n,b}_{\mathbf{k} = 0}\rangle\langle \phi^{n,b}_{\mathbf{k} = 0}|\right] \\
+ iA\left[|\phi^{n,a}_{\mathbf{k} = 0} \rangle\langle \phi^{n+1,b}_{\mathbf{k} = 0}| - |\phi^{n+1, b}_{\mathbf{k} = 0}\rangle\langle \phi^{n,a}_{\mathbf{k} = 0}|\right].
\end{multline}
An excerpt of the series of $(2\times2)$ Hamiltonians is
\begin{equation}
H^{\mathbf{k}=0}_{\text{F}} = \begin{pmatrix}
n\Omega & iA & 0 & 0\\
-iA & (n+1)\Omega & 0 & 0\\
0 & 0 & (n-1)\Omega & iA\\
0 & 0 & -iA & n\Omega
\end{pmatrix}.
\label{circh0}
\end{equation}
Out of the four eigenenergies of this matrix we are interested in
\begin{align}
\epsilon_{l}^{\pm} &= l\Omega \pm \frac{1}{2}\Delta,
\end{align}
with
\begin{align*}
\Delta &= \widetilde{\Omega}-\Omega\nonumber\\
\widetilde{\Omega} &= \sqrt{4A^2 + \Omega^2}.
\end{align*}
These two energies fulfill $\lim_{A\to0}\epsilon_{0}^{\pm}=0$, thus, we associate the first Brillouin zone, $l = 0$, from the Floquet solutions, to the solutions corresponding to graphene in the absence of an external field.
The corresponding eigenvectors  are
\begin{align}
|\phi_{l}^{+}\rangle &= \frac{1}{N}\left(2iA |\phi^{l-1, a}_{\mathbf{k} = 0}\rangle + (\Delta+2\Omega)|\phi^{l,b}_{\mathbf{k} = 0}\rangle\right)\\
|\phi_{l}^{-}\rangle &= \frac{1}{N}\left((\Delta+2\Omega)|\phi^{l,a}_{\mathbf{k} = 0}\rangle+2iA|\phi^{l+1,b}_{\mathbf{k} = 0}\rangle \right),
\end{align}
where $N=\sqrt{4A^2+(\Delta+2\Omega)^2}$. From here on and for the rest of the paper, we neglect the index $\mathbf{k}$ in the energies and vectors, unless we have to distinguish between $\mathbf{k}$ and $\mathbf{k+q}$.

Using these eigenvectors as a basis, we rewrite the full Floquet Hamiltonian \eqref{hamilfloquetcirc}
\begin{widetext}
\begin{align}
H_{\text{F}}=
\small
&\begin{pmatrix}[c|cc|cc|cc|c]
\ddots & \vdots & \vdots & \vdots & \vdots & \vdots & \vdots & \iddots\\[5pt]
\hline
\hdots & \epsilon_{n-1}^{+}& F_0k e^{i\Theta} &  F_1 k e^{i\Theta} & 0 & 0 & 0 & \hdots\\[5pt]
\hdots & F_0k e^{-i\Theta} & \epsilon_{n-1}^{-} & 0 & F_1^{*}k e^{i\Theta} & F_{2}k e^{i\Theta} & 0 & \hdots\\[5pt]
\hline
\hdots & F_{1}^{*}ke^{-i\Theta} & 0 & \epsilon_{n}^{+} & F_0 k e^{i\Theta} & F_{1}k e^{i\Theta} & 0 & \hdots\\[5pt]
\hdots & 0 & F_{1}ke^{-i\Theta} & F_{0}ke^{-i\Theta} & \epsilon_{n}^{-} & 0 & F_{1}^{*}ke^{i\Theta} & \hdots\\[5pt]
\hline
\hdots & 0 & F_{2}ke^{-i\Theta} & F_{1}^{*}k e^{-i\Theta} & 0 & \epsilon_{n+1}^{+} & F_{0}ke^{i\Theta}& \hdots\\[5pt]
\hdots & 0 & 0 & 0 & F_{1}ke^{-i\Theta} & F_{0}ke^{-i\Theta} & \epsilon_{n+1}^{-}& \hdots\\[5pt]
\hline
\iddots & \vdots & \vdots & \vdots & \vdots & \vdots & \vdots & \ddots
\end{pmatrix},
\label{HF0F1F2}
\end{align}
\end{widetext}
where $k=|\mathbf{k}|$, $\Theta=\arctan{k_y/k_x}$ and we introduced the three functions $F_0$, $F_1$ and $F_2$ that will form the basis for our approximations:
\begin{align}
F_0 &= \frac{(\Delta+2\Omega)^2}{4A^2+(\Delta+2\Omega)^2}\nonumber\\
F_1 &= \frac{2iA(\Delta+2\Omega)}{4A^2+(\Delta+2\Omega)^2}\nonumber\\
F_2 &= \frac{4A^2}{4A^2+(\Delta+2\Omega)^2}.
\label{functionsF0F1F2}
\end{align}
All three functions in Eq.\eqref{functionsF0F1F2} are functions of $A$ and $\Omega$.
For small $A/\Omega\ll1$, one finds that $F_0\approx1$, and $F_{1,2}\approx0$, however, $F_1$ increases linearly whereas $F_2$ increases quadratically with $A$. Note that for a two-level system driven by a linearly polarized field, the $n$th order Bessel function $J_{n}\left(A/\Omega\right)$ plays the role of the function $F_{0,1,2}$ presented here, and for a complete analysis one has to consider Bessel functions up to infinite order, see e.g. Ref.~[\onlinecite{plateroPhysRep04,hausingerPRA10}]. Here however, the complete information lies in $F_{0,1,2}$.
In the subsequent analysis, we will at first only consider the couplings given by $F_0$, and then include also the couplings given by $F_1$. We will neglect $F_2$ in general, which is valid for small $A/\Omega$.

\subsubsection{$F_0$--approximation}
\label{sectionF0}

A first approximation consists in neglecting both $F_1$ and $F_2$ and considering only $F_0$, which connects energies with the same photon number $n$. This approximation is valid for the calculation of many observables as far as the dimensionless quantity $A/\Omega\ll1$ -- i.e. the field intensity is small compared to the frequency -- and we are interested in excitations in the low energy sector, as we will see below, when we analyze the excitation spectrum and the generalized density of states.  The resulting Hamiltonian \eqref{HF0F1F2} is then block diagonal with building blocks $H_{\text{F}_0}^{n}$, where the matrix $H_{\text{F}_0}^{n}$ reads

\begin{align}
H_{\text{F}_0}^{n} =
\begin{pmatrix}
\epsilon_{n}^{+} & F_0ke^{i\Theta}\\
F_0ke^{-i\Theta} & \epsilon_{n}^{-}
\end{pmatrix}.
\end{align}
Its eigenvalues and eigenvectors are
\begin{align}
\epsilon_{l,\text{F}_0}^{\pm} &= l\Omega \pm \frac{1}{2}\sqrt{4F_0^2k^2+\Delta^2}
\label{quasiF0}
\end{align}
\begin{align}
|\chi_{l,\text{F}_0}^{+}\rangle &=
 \frac{1}{\sqrt{|\chi_a|^2+|\chi_b|^2}}\left(\chi_a|\phi_{l}^{+}\rangle + \chi_b|\phi_{l}^{-}\rangle\right)\nonumber\\
|\chi_{l,\text{F}_0}^{-}\rangle &=
 \frac{1}{\sqrt{|\chi_a|^2+|\chi_b|^2}}\left(\chi_{b}^{*}|\phi_{l}^{+}\rangle - \chi_{a}^{*}|\phi_{l}^{-}\rangle\right),
 \label{quasiF0vec}
\end{align}
where
\begin{align}
\chi_a &= 2F_{0}ke^{i\frac{\Theta}{2}}\\
\chi_b &= \left(\sqrt{4F_{0}^{2}k^2+\Delta^2}-\Delta\right)e^{-i\frac{\Theta}{2}}.
\label{chiachib}
\end{align}
The main virtue of this approximation is the fact that it captures the gap $\Delta$ produced at $\mathbf{k}=0$ by the ac electric field, giving an analytical expression for its magnitude,  $\Delta = \sqrt{4A^2+\Omega^2} - \Omega$, so this gap can be tuned by varying the field strength of the applied ac field, see also Refs.~[\onlinecite{aokiPRB09, calvoAPL11, zhouPRB11}].  We point out that an analogous phenomenon occurs in the optics of semiconductors in a strong THz-field: 
there the dynamical Franz-Keldysh effect\cite{anttiPRL96, nordstromPRL98} blue-shifts the conduction band edge (or, equivalently, the optical absorption edge) by the ponderomotive energy, which also depends quadratically on the ac-field amplitude.
This so called {\it $F_0$--approximation} neglects the coupling between Hamiltonians with a different number of photons, and is therefore not useful once we are interested in the anticrossings of the Floquet quasienergies for non-zero momentum.

\subsubsection{$F_1$--approximation}
\label{sectionF1}

In order to analyze higher order processes, we go one step further and take into account the coupling elements $F_1$, which capture the one-photon resonances, yielding a much more robust approximation for the Hamiltonian $H_{\text{F}}$ \eqref{HF0F1F2}.
At the resonances the relevant couplings are the ones between $\epsilon_{n-1}^+$ and $\epsilon_{n}^-$, $\epsilon_{n}^+$ and $\epsilon_{n+1}^-$ etc., see Eq.~\eqref{HF0F1F2}.
By applying the unitary matrix that diagonalizes $H_{\text{F}_0}^{n}$,
\begin{align}
U_{n}=\frac{1}{\sqrt{|\chi_a|^2+|\chi_b|^2}}\begin{pmatrix}
\chi_a & \chi_{b}^{*}\\
\chi_b & -\chi_{a}^{*}
\end{pmatrix}
\label{UF0F1}
\end{align}
we can construct a new effective infinite Hamiltonian which includes the features of the one--photon resonance, and which is again block diagonal, now mixing the sectors that differ in one photon in the {$F_0$--approximation}:
\begin{align}
H_{\text{F}_1}^{\text{eff},n} =
\begin{pmatrix}
\epsilon_{n-1, \text{F}_0}^{+} & \frac{2}{S_k}F_0F_1k^2e^{i\Theta} \\
\frac{2}{S_k}F_0F_1^{*}k^2e^{-i\Theta} & \epsilon_{n, \text{F}_0}^{-}
\end{pmatrix},
\end{align}
where $S_{k} = \sqrt{4F_0^2k^2+\Delta^2}$. The intensity of the coupling is proportional to $F_1$, as expected. Diagonalizing this Hamiltonian yields the following Floquet quasienergies:

\begin{widetext}
\begin{equation}
  \epsilon_{l,\text{F}_1}^{+}  = \left\{
  \begin{array}{l l}
    l\Omega + \frac{1}{2}\left(\Omega-\sqrt{\left(\Omega-S_{k}\right)^2 + \frac{16}{S_k^2}F_0^2|F_1|^2k^4}\right) & \quad \text{if $k<k_c$}\\
    (l+1)\Omega - \frac{1}{2}\left(\Omega-\sqrt{\left(\Omega-S_{k}\right)^2 + \frac{16}{S_k^2}F_0^2|F_1|^2k^4}\right) & \quad \text{if $k>k_c$}\\
  \end{array} \right.
  \label{quasiF1a}
\end{equation}
\begin{equation}
  \epsilon_{l,\text{F}_1}^{-}  = \left\{
  \begin{array}{l l}
    l\Omega - \frac{1}{2}\left(\Omega-\sqrt{\left(\Omega-S_{k}\right)^2 + \frac{16}{S_k^2}F_0^2|F_1|^2k^4}\right) & \quad \text{if $k<k_c$}\\
    (l-1)\Omega + \frac{1}{2}\left(\Omega-\sqrt{\left(\Omega-S_{k}\right)^2 + \frac{16}{S_k^2}F_0^2|F_1|^2k^4}\right) & \quad \text{if $k>k_c$},
  \end{array} \right.
  \label{quasiF1b}
\end{equation}
where  $k_c$ is the momentum at which the one--photon resonance takes place. For the Floquet eigenvectors, it is more convenient to write them in the basis that diagonalizes the Hamiltonian for $\mathbf{k} = 0$, reading
\begin{equation}
|\xi_{l,\text{F}_1}^{+}\rangle = \left\{
\begin{array}{l l}
\frac{1}{\sqrt{(|\chi_{a}|^2+|\chi_{b}|^2)(|\xi_{a}|^2+|\xi_{b}|^2)}}
\left[\xi_{a}\chi_{a}|\phi_{l}^{+}\rangle+\xi_{a}\chi_{b}|\phi_{l}^{-}\rangle - \xi_{b}\chi_{b}^{*}|\phi_{l+1}^{+}\rangle + \xi_{b}\chi_{a}^{*}|\phi_{l+1}^{-}\rangle\right]& \quad \text{if $k<k_c$}\\
\frac{1}{\sqrt{(|\chi_{a}|^2+|\chi_{b}|^2)(|\xi_{a}|^2+|\xi_{b}|^2)}}
\left[\xi_{b}^{*}\chi_{a}|\phi_{l}^{+}\rangle+\xi_{b}^{*}\chi_{b}|\phi_{l}^{-}\rangle + \xi_{a}^{*}\chi_{b}^{*}|\phi_{l+1}^{+}\rangle - \xi_{a}^{*}\chi_{a}^{*}|\phi_{l+1}^{-}\rangle\right]& \quad \text{if $k>k_c$}
\end{array} \right.
\label{quasiF1avec}
\end{equation}
\begin{equation}
|\xi_{l,\text{F}_1}^{-}\rangle = \left\{
\begin{array}{l l}
\frac{1}{\sqrt{(|\chi_{a}|^2+|\chi_{b}|^2)(|\xi_{a}|^2+|\xi_{b}|^2)}}
\left[\xi_{b}^{*}\chi_{a}|\phi_{l-1}^{+}\rangle+\xi_{b}^{*}\chi_{b}|\phi_{l-1}^{-}\rangle + \xi_{a}^{*}\chi_{b}^{*}|\phi_{l}^{+}\rangle - \xi_{a}^{*}\chi_{a}^{*}|\phi_{l}^{-}\rangle\right]& \quad \text{if $k<k_c$}\\
\frac{1}{\sqrt{(|\chi_{a}|^2+|\chi_{b}|^2)(|\xi_{a}|^2+|\xi_{b}|^2)}}
\left[\xi_{a}\chi_{a}|\phi_{l-1}^{+}\rangle+\xi_{a}\chi_{b}|\phi_{l-1}^{-}\rangle - \xi_{b}\chi_{b}^{*}|\phi_{l}^{+}\rangle + \xi_{b}\chi_{a}^{*}|\phi_{l}^{-}\rangle\right]& \quad \text{if $k>k_c$},
\end{array} \right.
\label{quasiF1bvec}
\end{equation}
\end{widetext}
where we have introduced
\begin{align}
\xi_a &= \left(\left(\Omega-S_k\right)+\sqrt{\left(\Omega-S_k\right)^2+\frac{16}{S_k^2}F_0^2|F_1|^2k^4}\right)e^{\frac{i}{2}\Theta}\\
\xi_b &= \frac{4}{S_k}F_0F_1^*k^2e^{-\frac{i}{2}\Theta}.
\end{align}
The {\it $F_1$--approximation} captures the gap at $\mathbf{k} = 0$ as well as the first resonance. The latter gives rise to the opening of new gaps, whose expression can be obtained analytically in this approximation, yielding
\begin{align}
\Delta_{\text{F}_1} = \sqrt{\left(S_{k_c}-\Omega\right)^2 + \frac{16}{S_{k_c}^2}F_0^2F_1^2k_{c}^{4}}.
\label{gapfirstresonance}
\end{align}
For a frequency $\Omega\approx150$meV in the mid-infrared regime, and field intensity $E_{0}\approx4.8$MV/m (so that $A/\Omega=0.1$), the size of the two gaps would be $\Delta\approx3$meV, $\Delta_{\mathrm{F}_1}\approx15$meV.

\subsection{Single particle properties of the Hamiltonian derived from the analytical approximations}
\label{sec2subsec3}

\begin{figure}[h]
\begin{center}
\includegraphics[width=3.4in,clip] {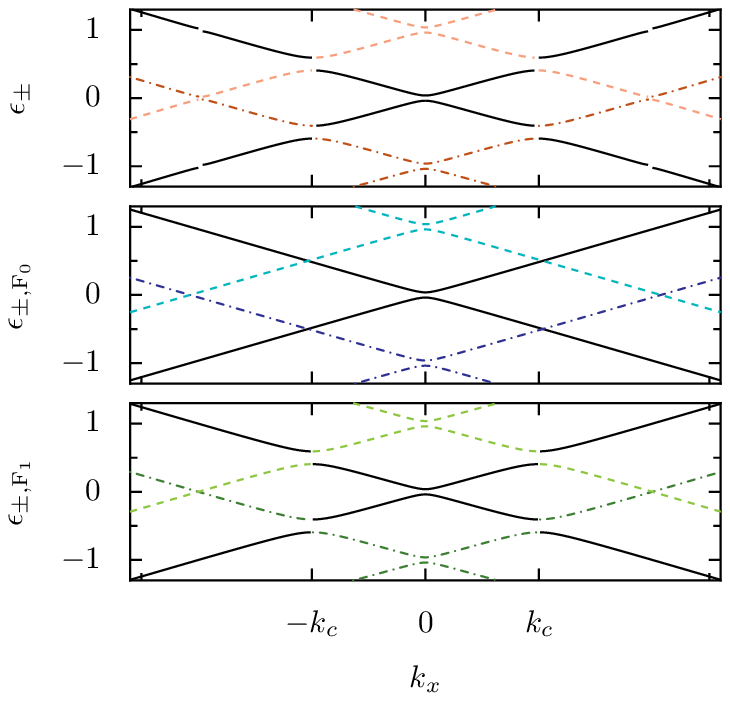}
\end{center}
\caption{\label{fig:quasikxcirc}\small
Quasienergy spectrum as a function of $k_x$ for $k_y=0$. The solid lines represent the $l=0$ band, the dashed lines the $l=1$, and the dashdotted lines the $l=-1$ sideband.
{\it Upper panel}:
The full numerical result of the quasienergies.
{\it Middle panel}:
The quasienergies for the  zero--photon approximation $F_0$.
{\it Lower panel}:
The quasienergies for the one--photon approximation $F_1$.
Parameters:  $k_y=0$, $A = 0.2$, $\Omega=1$.}
\end{figure}

We next consider the quasienergy spectrum for circularly polarized field, both the full numerical result and the analytical approximation derived in the previous sections.
With increasing field strength, zero--photon, one--photon, two--photon and higher order resonances appear.
In Fig.~\ref{fig:quasikxcirc} we compare the numerical (upper panel) and the analytical results (middle and lower panels) for the quasienergy spectrum as a function of the wavevector $k_x$ and for weak fields.
In the middle panel we plot the $F_0$--approximation, which reproduces very well the gap at $k_x=0$, but no other features induced by the ac field show up.
The lower panel shows the results for the $F_1$--approximation, which in addition to the $F_0$ result captures nicely the one--photon resonance at $k_c\approx\pm0.5$.

The analytical approximations can be tested by computing the density of states (DOS), which was already analyzed numerically using the full Floquet Hamiltonian by Oka {\it et al.},\cite{aokiPRB09} Calvo {\it et al.},\cite{calvoAPL11} and Zhou {\it et al.}.\cite{zhouPRB11}  The generalized (time averaged) density of states can be calculated as:

\begin{equation}
D(\omega) = 4\sum_{\mathbf{k},\sigma}\delta(\omega - \epsilon_{\mathbf{k}, \sigma,  0}),
\end{equation}
where the quasienergies are those defined in the first Brillouin zone of the Floquet spectrum. In the $F_0$-- approximation,
$\epsilon_{\mathbf{k}, \sigma, l=0} = \epsilon_{0,\text{F}_0}^{\pm}$, see Eq.~\eqref{quasiF0}, and the density of states can be calculated analytically yielding
\begin{align}
D_{\text{F}_0}(\omega) &=
\frac{2}{\pi F_0^2}|\omega|\Theta\left(|\omega|-\frac{\Delta}{2}\right).
\end{align}
Notice the presence of the gap at zero energy in the density of states.

\begin{figure}[h]
\begin{center}
\includegraphics[width=3.2in,clip] {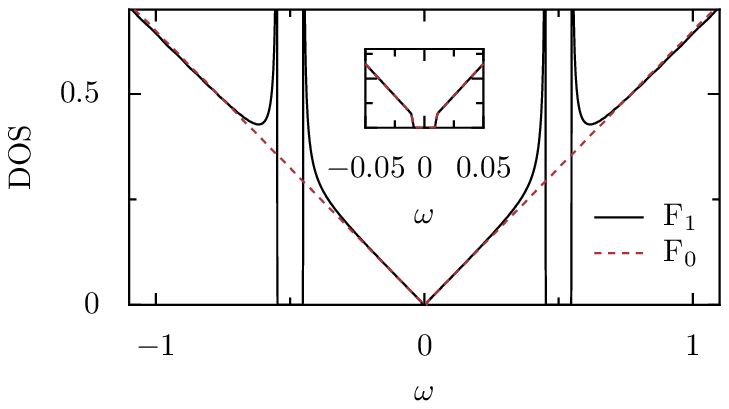}
\end{center}
\caption{\label{fig:doscircF0F1}\small
Density of states versus energy for the analytical approximations $F_0$ and $F_1$ (upper panel) and considering the full Floquet Hamiltonian Eq.~\eqref{hamilfloquetcirc} (lower panel).
The $F_0$--approximation reproduces the gap given by $\Delta$, see text.
For the one--photon resonance in the $F_1$--approximation, the gap at $\omega=0.5$ is reproduced,
where the field couples modes with $n$ and $n+1$ photons. 
For the same field strength as considered in the $F_0$ and $F_1$ approximations, the full numerical density of states is identical up to an additional resonance at $\omega\approx1$, which is due to two--photon processes. 
In the inset, the region around the gap $\omega=0$ is blown up for better visibility.
Parameters:  $A=0.1$, $\Omega = 1$.}
\end{figure}

The simplicity of this zero-photon approximation allows for analytical computations of many physical quantities, something that no longer happens in general  in the one--photon approximation, for which we have to resort to numerical calculations in most of the cases. For the generalized density of states, by using the analytical quasienergies, Eqs.~\eqref{quasiF1a} and \eqref{quasiF1b}, the results of both the $F_0$-- and $F_1$--approximation are plotted in Fig.~\ref{fig:doscircF0F1} (upper panel). As a comparison, the lower panel of Fig.~\ref{fig:doscircF0F1} shows the density of states calculated numerically by diagonalizing the full Floquet Hamiltonian \eqref{hamilfloquetcirc}. Once again, notice that the $F_0$--approximation works very well for energies of the order of the first gap, while in order to study the first resonance the $F_1$--approximation excels quite well. 
Higher resonances -- visible in the numerical result for the density of states at around $\omega\approx1$, are almost negligible for the field strength we are considering here.\\
Although the density of states has already been achieved numerically by various authors,\cite{aokiPRB09, calvoAPL11, zhouPRB11} the developed approximations are useful in order to both obtain analytical results when this is possible or at least simplify the numerical complexity of the problem, as it happens when we use the one--photon approximations. In principle, these approximations are valid for arbitrary $\mathbf{k}$, as long as $A/\Omega\ll1$. The results for the density of states show that in both analytical approximations and in the full numerical case the gaps remain stable independently of the range of integration in the momentum included in the density of states, i.e., the inclusion of higher momentum states does not close the gaps in our case, contrary to what was found in Ref.~[\onlinecite{zhouPRB11}].

\section{Single and many-particle excitations in graphene in a circularly polarized ac electric field}
\label{sec3}

\subsection{Electron interactions and the formula for the dynamical polarizability}
\label{sec3subsec1}
So far we have analyzed the single particle properties of the Hamiltonian. However, a full description of electron excitations in graphene requires to understand the role of electron-electron interactions in the system. The Hamiltonian of the interacting system in the presence of an ac field reads now, in second quantization,
\begin{equation}
H(t) = v_{\mathrm{F}}\sum_{\mathbf{k}} \Psi_{\mathbf{k}}^{\dagger}\boldsymbol\sigma \cdot (\mathbf{k} - e \mathbf{A}(t))\Psi_{\mathbf{k}} + \sum_{\mathbf{q}} v_{\mathbf{q}} n_{\mathbf{q}}^{\dagger} n_{\mathbf{q}},
\label{eq:graphene_hamilV}
\end{equation} 
where $v_{\mathbf{q}}= 2\pi e^2/\epsilon_0 q$ is the 2D unscreened Coulomb interaction. In the absence of the external ac field, this Hamiltonian has been extensively studied (for a review, see ref.~\onlinecite{castronetoRMP11}). For doped samples of graphene, the Coulomb interaction becomes screened, yielding a system whose low-energy excitations around the Fermi surface are barely interacting, i.e., electrons in graphene behave as a Fermi liquid. Moreover, a collective excitation, a plasmon, exists.\cite{dassarmaPRB07, wunschNJoP06} This is no longer true when the level of doping is zero or very small, where the role of interactions is controversial due to the singular nature of the Dirac point, where screening is uneffective.\cite{gonzalezNPB94, mishchenkopolPRL08}
In order to understand the effect of interactions between electrons when an external ac field is applied, we will compute the dynamical polarizability, which tells us about the response of the system to probes that couple to the electric charge. This function will yield information about the full spectrum of electron interactions of the system, that contains both single particle and collective excitations. The dynamical polarizability of graphene in the presence of an ac electric field shows particular features that differ from its counterpart, the two dimensional electron gas, as well as from the one derived for graphene in the absence of the field.\cite{shungPRB86} 
The detailed derivation of the polarizability function is given in the Appendix, whereas here we present the final result:
\begin{eqnarray}
\Pi(\mathbf{q},\omega) =  \sum_{\sigma,\sigma'}\sum_{\mathbf{k}}\sum_{l}\frac{f_{\mathbf{k},\sigma}-f_{\mathbf{k+q},\sigma'}}{\omega - \epsilon_{\mathbf{k+q},\sigma',l}+\epsilon_{\mathbf{k},\sigma,0}+i\eta}\nonumber\\
\times \sum_{n}|\phi_{\mathbf{k+q},\sigma',l}^{n,a,*}\phi_{\mathbf{k},\sigma,0}^{n,a}+\phi_{\mathbf{k+q},\sigma',l}^{n,b,*}\phi_{\mathbf{k},\sigma,0}^{n,b}|^2
\label{polarizability}
\end{eqnarray}
The index $n$ stands for the Fourier component of a solution in the sideband $l$ of the infinite Floquet Hamiltonian. The summation over $n$ constitutes the scalar product of the solution $|\phi_{\mathbf{k+q},\sigma',l}\rangle$ with $|\phi_{\mathbf{k},\sigma,0}\rangle$, where we have used the fact that solutions belonging to the $l$th Brilloun zone in the Floquet spectrum are those of the first Brilloun zone shifted by $l$ units, see Eqs.~\eqref{transformation1} and \eqref{transformation2}. The solution $l=0$ is the one which fulfills the condition that at $A\rightarrow0$, $\Pi(\mathbf{q},\omega)$ becomes the polarizability for an isolated graphene sheet. 
Mathematically, it is important to notice that the analytical properties of the dynamical polarizability do not change in the presence of the external ac field. It is a complex function, analytical in the upper half plane,  whose real and imaginary parts are not independent, but related via the Kramers-Kronig relations, see e.g. ref.~\onlinecite{jackson}. The latter can also be seen as a consequence of the causality in the response of the system to the external probe.

The polarizability is written in terms of the single particle excitations of the system, as  in conventional linear response theory. In the presence of the ac electric field, there is an infinite set of single particle excitations that differ between them in the relative number of photons. Once the external field is switched on, the system is no longer isolated, and the field can pump or extract energy into the system in the form of photons of frequency $\Omega$. Therefore, the polarizability can be seen as a linear combination of polarizabilities, each describing excitations in which the number of photons in the system changes by a certain integer number $l$. Or in other words, the response of the system to an external probe of energy $\omega$ and momentum $q$ can arise from excitations in which no extra photons are introduced in the system, as it happens in the absence of the field, or also in which a given number $l$ of photons is introduced or extracted from it.
Similar results as those shown here have been derived in the context of low-dimensional semiconductors,\cite{anttiPRL99} and the 2DEG.\cite{chinodynscreeningPRB02} The latter is usually the benchmark to compare the results derived for graphene, and therefore we include here the formula for the ease of comparison:
\begin{equation}
\Pi^{2{\rm DEG}}_{ac}(\mathbf{q},\omega) = \sum_{\mathbf{k}}\sum_{l}\frac{f_{\mathbf{k}}-f_{\mathbf{k+q}}}{\omega - \epsilon_{\mathbf{k+q},l}+\epsilon_{\mathbf{k},0} + i\eta}
\end{equation}
Note in this expression that the effects due to the ac electric field  appear in two different places: (i) the index $l$ of the sideband reflecting a change in the number of photons, and (ii) as a modification of the single particle excitations in $\epsilon_{\mathbf{k},\sigma,l}$.
The situation is more complicated in the case of graphene, Eq.~(\ref{polarizability}), where also
a momentum dependent overlap term between the excitations with different momentum must be included. This overlap is reflected in the polarizability via the scalar product between quasieigenstates, and it is in turn a consequence of the existence of the pseudospin in graphene. The effect of the the electric field on the electronic system can be understood in terms of transfer of spectral weight, as we shall show in the next section. As the electric field is switched on, the spectral weight is reorganized, although in a way that still preserves the conservation rule imposed by the $f$-sum rule, that was derived and analyzed in the context of low-energy graphene by Sabio {\it et al.}\cite{sabiofsumPRB08}. 

As a last remark, since we are dealing with a system in which a polarized ac electric field is already present, one has to wonder about the possible influence of the polarization of the external probe to which the system responds. In fact, since we are analyzing the dynamical polarizability, which arises from the coupling between the electronic density and the potential induced by the external probe, the response of the system in the linear regime is insensitive to the polarization of the probe field. In order to see a response that depends on the polarization of the probe, we would have to analyze the response of the electronic current, which couples to the electric field, and whose linear response function is the conductivity. Notice, however, that the response function itself will not be altered due to this polarization, since in linear response it only depends on the properties of the system in the absence of the external probe by virtue of the fluctuation-dissipation theorem.

\subsection{Analytic approximations for the dynamical polarizability}
\label{sec3subsec2}
We now evaluate the dynamical polarizability (\ref{polarizability}) using the analytic approximations developed in Sect.~\ref{sec2subsec2}.
We first consider the imaginary part of (\ref{polarizability}). This yields the response of the non-interacting system to an external probe of energy $\omega$ and momentum $q$, and is a building block for the Random Phase Approximation (RPA).
RPA is known to work well for doped graphene,\cite{castronetoRMP11} where Landau's theory of the Fermi liquid provides a good description of the low energy excitations. In the case of undoped graphene, the issue is more complicated due to the lack of screening near the Dirac point.\cite{mishchenkopolPRL08} In our case, due to the opening of a gap at $\mathbf{k}=0$, we expect RPA to be sufficient to describe the main features of the response of the system once electron-electron interactions are taken into account.

\subsubsection{$F_0$--approximation}

\begin{figure}[h!]
\begin{center}
\includegraphics[width=3.4in,clip] {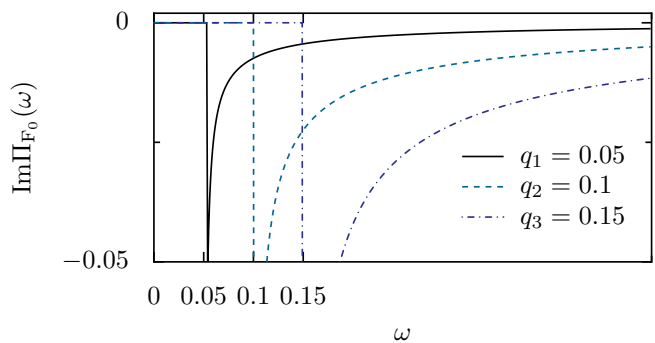}
\caption{\label{figimpolF0}\small
Imaginary part of the polarizability in the $F_0$--approximation as a function of $\omega$ and for different values of $q$, Eq.~\eqref{impolF0}. For $\omega<\sqrt{F_{0}^2q^2+\Delta^2}$, no excitations are possible, as compared to free graphene, where no excitations are possible for $\omega<q$. Note that only for small $q = 0.05$, the effect of the gap is visible as a shift of the divergence away from $\omega=q$.
Parameters: $A=0.1$, $\Omega=1$.}
\end{center}
\end{figure}

\begin{figure}[h!]
\begin{center}
\includegraphics[width=3.4in,clip] {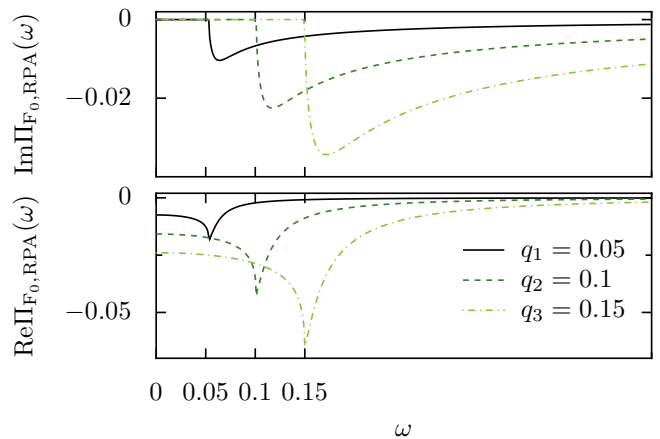}
\caption{\label{figRPAF0}\small
Imaginary and real part of the polarizability in the $F_0$--approximation as a function of $\omega$ for different $q$ in the RPA. The divergence in the imaginary part of the polarizability has disappeared, but the real part has not developed a resonance, which would be the signature of the existence of collective electronic excitations.
Parameters: $A=0.1$, $\Omega=1$.}
\end{center}
\end{figure}

In the $F_0$--approximation a gap opens up at zero momentum, so it can be used for a qualitative description of the response to an external ac probe. We set $T=0$, and for the moment we will restrict ourselves to undoped graphene. In this case, the dynamical polarization involves only the term that accounts for transitions between the Floquet bands $(0, -)$ and $(0, +)$, i.e., transitions in which the number of photons is conserved:
\begin{align}
\Pi_{\text{F}_0}(\mathbf{q},\omega) = \sum_{\mathbf{k}}
\frac{|\langle\chi_{\mathbf{k+q},0}^{+}|\chi_{\mathbf{k},0}^{-}\rangle|^2}{\omega-\epsilon_{\mathbf{k+q},0}^{+}+\epsilon_{\mathbf{k},0}^{-}+i\eta},
\label{polF0}
\end{align}
where $|\chi_{\mathbf{k},0}^{\pm}\rangle$ are the eigenvectors from the $F_0$--approximation, see Eq.~\eqref{quasiF0vec}. We find
\begin{multline}
\operatorname{Im}\Pi_{\text{F}_0}(\mathbf{q},\omega) = -\frac{1}{4}\frac{F_{0}^2q^2}{\sqrt{\omega^2-F_0^2q^2}}\\
\times\left(1+\frac{\Delta^2}{\omega^2-F_0^2q^2}\right)\Theta\left(\omega^2-F_0^2q^2-\Delta^2\right),
\label{impolF0}
\end{multline}
with $\Delta = \sqrt{4A^2+\Omega^2}-\Omega$ being the gap opened at the first resonance. This is actually the dynamical polarizability for a ``gapped graphene";  now the gap is due to the presence of the circularly polarized ac field. Gapped graphene has been studied extensively by Pyatkovskiy,\cite{russeJoPCM09} who also derived analytical expressions for the real part of the polarization, for both doped and undoped graphene.
We show the imaginary part of the polarizability in the $F_0$--approximation in Fig.~\ref{figimpolF0}. For a given momentum of the external probe, the energy threshold required to produce single particle excitations is increased due to the existence of the gap, being located now at $\omega = \sqrt{F_0^2 q^2 + \Delta^2}$. This yields a rearrangement of the spectral weight of the excitations, which might allow for the existence of more complex excitations in the spectrum of the interacting system. We investigate this question within RPA.
The polarizability in the RPA is
\begin{equation}
\Pi_{{\rm RPA}}(\mathbf{q}, \omega) = \frac{\Pi_{0}(\mathbf{q}, \omega)}{1 - v_q \Pi_{0}(\mathbf{q}, \omega)},
\label{RPAPol}
\end{equation}
where the denominator is the dielectric function in RPA with $v_q$ being the 2D unscreened Coulomb potential. In order to have long-lived collective excitations, i.e. plasmons, the dielectric function must vanish at certain points $\omega_{p}(q)$, which leads to the conditions $v_q \operatorname{Re} \Pi_{0}(\mathbf{q}, \omega) = 1$ and $\operatorname{Im}\Pi_{0}(\mathbf{q}, \omega) = 0$.
In Fig.~\ref{figRPAF0} the imaginary and real part of the polarizability in the RPA are plotted, where we use in Eq.~\eqref{RPAPol} $\Pi_{0}(\mathbf{q},\omega)=\Pi_{\text{F}_0}(\mathbf{q},\omega)$ (Eq.~\eqref{polF0}). It can be seen that the divergence in the threshold of excitations found for the non-interacting polarizability has disappeared, a feature that is also observed in graphene in the absence of ac fields. However, the real part of the polarizability does not develop a resonance, leading to the absence of plasmons, at least within the RPA of the $F_0$--approximation.
In order to test if this is still true when higher order photon resonances are included, we analyze the $F_1$--approximation in the next section.

Before that, let us analyze the case of doped graphene. As already shown in Ref.~[\onlinecite{russeJoPCM09}] for the case of gapped graphene,  the plasmon already present in the system without ac fields is still robust once a weak field is introduced. The main effect of the external ac field is to modify the plasmon dispersion:
\begin{align}
\omega_{p}^{\text{F}_0}(q) = \sqrt{\frac{g N_s N_v F_{0} q \mu}{2}\left(1-\frac{\Delta^2}{4\mu^2}\right)},
\end{align}
where $g = e^2/v_F \epsilon_0 \hbar$ is the fine structure constant of graphene.
The correction affects the plasmon frequency $\omega_0 = \sqrt{\frac{g N_s N_v F_{0} \mu}{2}(1-\frac{\Delta^2}{4\mu^2})}$, but not the dependence on momentum, which still follows the law $\omega_{p}^{\text{F}_0}(q) \propto \sqrt{q}$. The plasmon frequency is diminished due to the effect of the external ac field, since $F_0 < 1$ and $1- \Delta^2/(4 \mu^2) < 1$. The correction coming from the factor $F_0$ is essentially due to the renormalization of the Fermi velocity due to the ac field. The second correction depends on the relation of the chemical potential to the gap at zero momentum, and it is maximal for a chemical potential below $\Delta / 2$, where the plasmon is completely suppressed since no electrons are populating the upper Dirac cone. For chemical potentials above this value, the correction tends to be smaller, being almost negligible for $\mu \gg \Delta/2$.
We point out that similar results hold in a quite different context, that of graphene anti-dot lattices,\cite{anttiPRL08, anttiPRB11} where it was found that in the limit of low doping, gapped graphene models reproduce very well the plasmon dispersion of the anti-dot lattice.

In short, the results from the $F_0$--approximation yield a similar picture to that of graphene in the absence of an external field, and the main effect of the ac field is a renormalization of the single and many-particle spectrum, with a shift of the threshold for excitations due to the gap at zero momentum.\\

\subsubsection{$F_1$--approximation}
The $F_1$--approximation
accounts for non--zero--photon processes, neglected in the $F_0$--approximation.
It is not fully analytically tractable in the calculation of many observables.
In the $F_1$--approximation using Eqs.~\eqref{quasiF1a}-\eqref{quasiF1bvec} the polarizability for undoped graphene at $T=0$ becomes
\begin{widetext}
\begin{align}
\Pi_{F_{1}}(\mathbf{q},\omega) = \sum_{\mathbf{k}}&\frac{|\langle\xi_{\mathbf{k+q},0}^{+}|\xi_{\mathbf{k},0}^{-}\rangle|^2}{\omega-\epsilon_{\mathbf{k+q},0}^{+}+\epsilon_{\mathbf{k},0}^{-}+i\eta}\nonumber\\
+&|\langle\xi_{\mathbf{k+q},-1}^{+}|\xi_{\mathbf{k},0}^{-}\rangle|^2\left(\frac{1}{\omega-\epsilon_{\mathbf{k+q},-1}^{+}+\epsilon_{\mathbf{k},0}^{-}+i\eta}-\frac{1}{\omega-\epsilon_{\mathbf{k+q},1}^{-}+\epsilon_{\mathbf{k},0}^{+}+i\eta}\right)\nonumber\\
+&|\langle\xi_{\mathbf{k+q},-2}^{+}|\xi_{\mathbf{k},0}^{-}\rangle|^2\left(\frac{1}{\omega-\epsilon_{\mathbf{k+q},-2}^{+}+\epsilon_{\mathbf{k},0}^{-}+i\eta}-\frac{1}{\omega-\epsilon_{\mathbf{k+q},2}^{-}+\epsilon_{\mathbf{k},0}^{+}+i\eta}\right)
\label{polF1}
\end{align}
\end{widetext}
Here, there are three different contributions to the polarizability, coming from the Floquet bands $l=0$, $l=\pm1$ and $l=\pm2$, i.e., from excitations that involve the exchange of up to two photons from the external field. However, for the electric fields in which this approximation holds, the contribution from the $l=\pm2$ components is essentially negligible, and therefore only zero and one--photon processes will be considered.
In what follows we evaluate (\ref{polF1}) numerically, first integrating the imaginary part and then computing the real part via the Kramers Kronig relations. 

\begin{figure}[h]
\includegraphics[width=3.4in,clip] {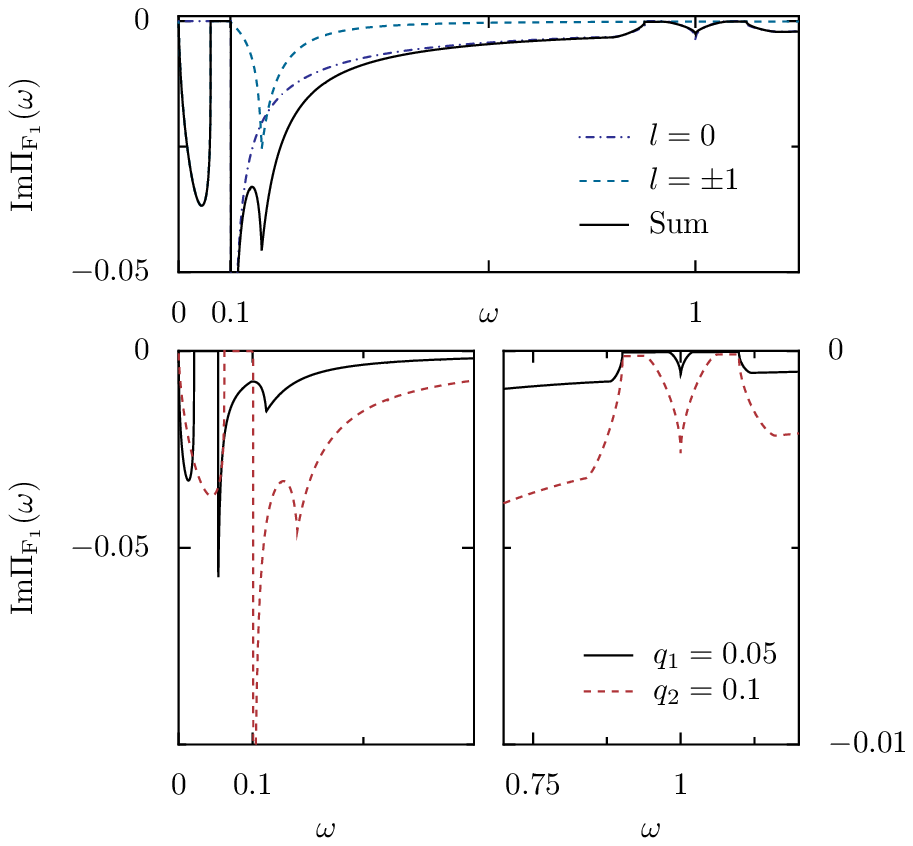}
\caption{\label{impol}\small
Imaginary part of the polarizability $\Pi_{\text{F}_1}$ as a function of $\omega$.
{\it Upper panel}: $q=0.1$; the components $l=0,\pm1$ of the polarizability and their sum are shown, see Eq.~\ref{polF1}.
{\it Lower panels}: Imaginary part of the total polarizability as a function of $\omega$ for two different $q$: $q_1 = 0.05$, $q_2 = 0.1$. Left plot for $\omega<0.4$, right plot for $\omega>0.7$.
Parameters: $A=0.1$, $\Omega=1$.
}
\end{figure}

Figure \ref{impol} shows the imaginary part of the polarizability $\Pi_{F_{1}}$ for fixed $q$ as a function of $\omega$. In the upper panel, the components $l=0$ and $l=\pm1$ and their sum are shown for $q = 0.1$, in order to illustrate where its structure comes from. The two lower panels represent the total polarizability for two different wavevectors $q$, divided into two regions for better visibility of the different features.
Several new features emerge from the $F_1$--approximation.
As shown in Fig.~\ref{impol}, at the level of zero--photon processes, there is a gap at zero momentum, already captured in the $F_0$--approximation.
In addition, gaps appear at higher momenta, where the first anticrossing of Floquet sidebands occurs (see Fig.~\ref{fig:quasikxcirc}). For a sufficiently small $q$ this second gap translates into two small gaps in the single particle excitation spectrum around $\omega\approx 1$, which are eventually closed for higher momenta, as shown in Fig.~\ref{impol} (lower panel, right plot). The first of those gaps, for $0.9<\omega<1$ , is due to the fact that for electrons from the lower cone of graphene no states are available in the upper band for those values of $q$ and $\omega$ due to the anticrossing of Floquet sidebands.
The second one, at $1<\omega<1.1$, is due to the lack of states in the lower cone in the region where this anticrossing occurs with lower Floquet sidebands.

The most important new features of the response of the system, however, come from the contribution of one--photon processes, in which transitions from the $l = 0$ to the $l= \pm 1$ sidebands are taken into account. New single particle excitations appear below $\omega = \sqrt{F_0^2q^2 + \Delta^2}$, leaving only a small region of energies where no excitations are found, a region which again is closed for sufficiently large $q$ (dashed line for $q=0.1$).

\begin{figure}[h]
\includegraphics[width=3.4in,clip] {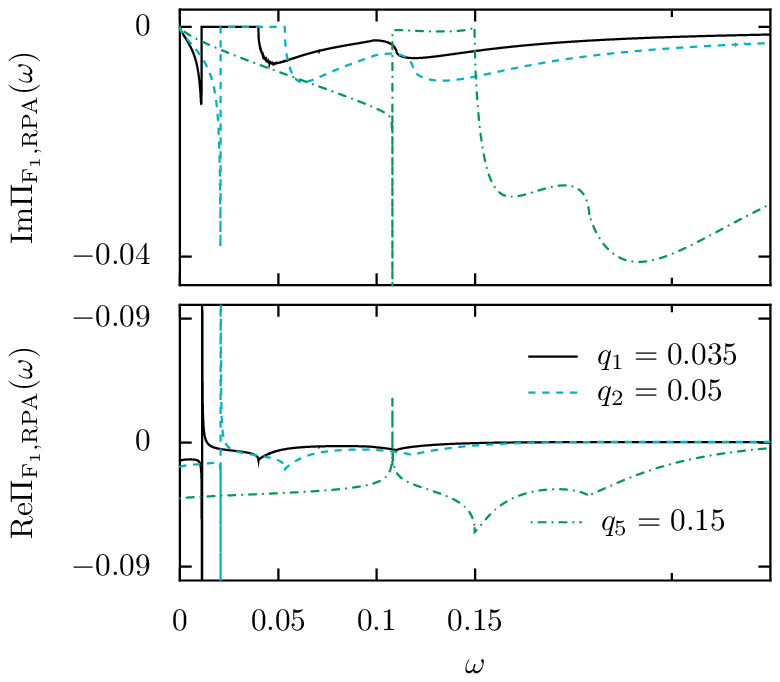}
\caption{\label{imRPAreRPA}\small
Imaginary and real part of the polarizability $\Pi_{\text{F}_1}$ in the RPA approximation. Notice the resonance in the real part for small momenta $q_1=0.035$ and $q_2=0.05$ (lower panel), where no excitations in the imaginary part exist (upper panel), pointing to the existence of collective excitations.
Parameters: $A=0.1$, $\Omega=1$.
}
\end{figure}

One--photon processes introduce new single particle excitations into the response of the system, and we next examine the effect of these processes on the collective excitations of the system. The RPA polarizability $\Pi_{\text{F}_1,\text{RPA}}$ is shown in Fig.~\ref{imRPAreRPA} for different values of the external momentum $q$. One--photon processes have an important effect on the response of the interacting system, allowing for the existence of collective excitations for small enough momentum, see curves for $q_1=0.035$, $q_2=0.05$ in Fig.~\ref{imRPAreRPA}. For those, the plasmon conditions are fulfilled, which is reflected in the development of a resonance in the real part of the RPA polarizability. For an external momentum of $q = 0.05$ e.g., the resonance is located at $\omega \simeq 0.021$, which is already in the region where the imaginary part of the polarizability is zero, allowing for an undamped plasmon. It is important to remark that this plasmon becomes unstable in two different scenarios. (i) For large enough momentum of the external probe, where the resonance is weakened and it occurs in a region where single particle excitations exist, so the plasmon can decay into those excitations, see $q_3=0.15$ in Fig.~\ref{imRPAreRPA}. (ii) When two--photon processes are considered, there is no longer a region where the imaginary part of the polarizability is zero. For weak fields, however, these processes are negligible and their effect on the plasmon should also essentially be irrelevant. However, this suggests that as we increase the intensity of the electric field, and higher order photon processes are important, there is no region of momenta $q$ in which the plasmon is stable.

\begin{figure}[h]
\includegraphics[width=3.4in,clip] {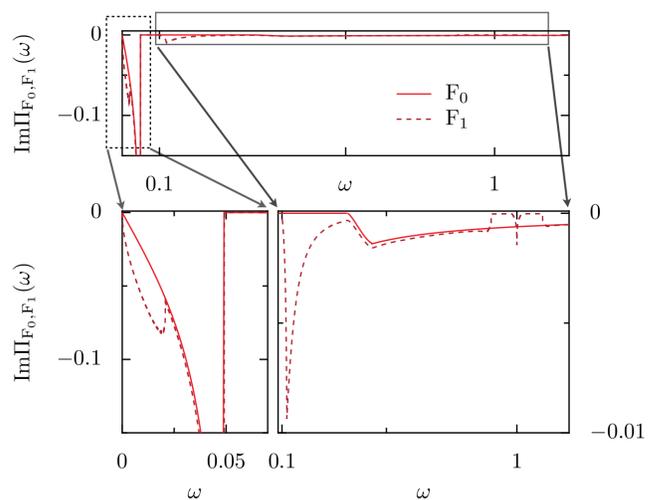}
\caption{\label{impoldoped}\small
Imaginary part of the polarizability for doped graphene as a function of $\omega$ for $q=0.05$.
Both the $F_{0}$- (solid red) and the $F_{1}$-approximation (dashed dark red) are shown.
In the lower panels, the upper plot is split in two parts in order to better visualize the different regions of $\omega$.
Parameters: $A=0.1$, $\Omega=1$, $\mu=0.2$.
}
\end{figure}

\begin{figure}[h]
\includegraphics[width=3.4in,clip] {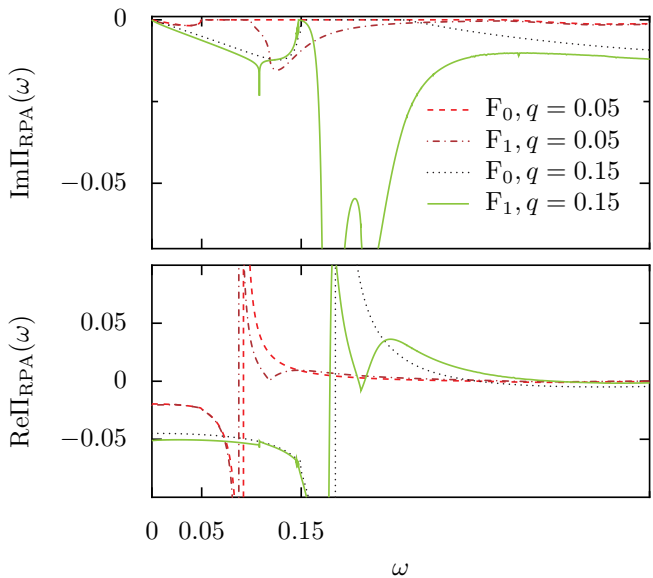}
\caption{\label{impoldopedRPA}\small
Polarizability in the RPA approximation for doped graphene in the $F_1$--approximation. {\it Upper panel}: imaginary part. {\it Lower panel}: real part. The results are shown for two different momenta $q = 0.05$ and $q = 0.15$. In both figures the results are compared with the $F_0$--approximation. Notice the existence, in both approximations, of the resonance in the real part of the polarizability, signalizing the existence of a collective excitation. For large momentum $q = 0.15$, however, the imaginary part shows no gap and therefore the plasmon can decay into single particle excitations.
Parameters: $A=0.1$, $\Omega=1$, $\mu=0.2$.
}
\end{figure}

For doped graphene, the $F_1$--approximation introduces similar features as those described for undoped graphene, see Fig.~\ref{impoldoped}.
The effect of the anticrossing of Floquet sidebands is to induce gaps in the response of the system for $\omega \sim 1$, and processes including the exchange of one photon give rise to new excitations for small energies. In order to quantify the effect of these processes, in this figure the polarizability is compared to the one for doped graphene, using the $F_0$--approximation, where only zero--photon processes are considered. The RPA response of the interacting system is shown for a couple of representative external momenta in Fig.~\ref{impoldopedRPA}. As it happened in the undoped case, for small external momentum ($q=0.05$ in Fig.~\ref{impoldopedRPA}) there is a resonance in the real part, signaling the existence of a plasmon, which again has a renormalized dispersion relation due to the effect of the external ac field.
However, non--zero--photon processes are responsible again for the appearance of low-energy excitations that tend to make the plasmon unstable for large enough momenta $q$ ($q=0.15$ in Fig.~\ref{impoldopedRPA}) and for larger intensities of the field, as discussed for the undoped case. These momenta $q$ for which plasmons become unstable are still lower than those for which the plasmon is damped in graphene without ac field.

Summarizing, the inclusion of non--zero--photon processes is crucial in order to capture the physics of the response of graphene to an external probe in the presence of a weak ac field. This is due to the appearance of excitations in the low energy spectrum of the system, not included in the $F_0$--approximation, that allow for the existence of collective excitations in undoped graphene, but make those plasmons unstable for smaller momenta than their counterparts in graphene with no ac fields.


\section{Conclusions}

In this work we have analyzed the properties of graphene under an external circularly polarized ac field in the weak field regime. We have developed analytical approximations to the Hamiltonian, the so called $F_0$ (see section \ref{sectionF0}) and $F_1$--approximations (see section \ref{sectionF1}), that allow a certain analytical tractability of many relevant objects. The $F_0$--approximation includes only zero--photon excitations in the system, and is useful to calculate certain observables in the low energy sector. The $F_1$--approximation includes higher order photon processes, allowing for the analysis of a wider range of observables and a larger energy sector. However, it requires in many cases numerical calculations to extract the observables.

Special emphasis has been put on the calculation of the polarizability of the system, which can be used to analyze the spectra of single and many-particle excitations of the system. We have derived a general expression for the polarizability of graphene in the presence of an ac electric field, which we have analyzed in the context of the $F_0$ and $F_1$--approximations. While the former allows for analytical expressions, and captures well the effect of the zero-momentum gap in the system, it misses the non--zero--photon processes, that are  captured by the $F_1$--approximation, and in turn are responsible for the emergence of collective excitations even for undoped graphene, as far as the Random Phase Approximation remains valid. However, it also points out that these collective excitations are less stable when compared to graphene with no external ac field:  for large enough external momenta and ac field intensities, these excitations become damped and acquire a finite lifetime.

 We have shown that circularly polarized ac fields can be used to modify the properties of graphene in several ways: (i) They open up gaps at zero momentum that can be exploited in practical applications, (ii) They permit the existence of plasmons (in both undoped and doped graphene), (ii) The plasmon frequency is tunable with the external field,  and, finally, (iv) For large enough fields the plasmons become unstable. Moreover, we have developed and tested analytical tools to analyze theoretically the behavior of graphene in the presence of ac electric fields, which should be useful in future works in this field.

\begin{acknowledgments}
The Center for Nanostructured Graphene is sponsored by the Danish National Research Foundation.

We would like to thank J. Sabio for fruitful discussions and T. Stauber for helpful comments.
We are grateful to Prof. M. W. Wu for communicating his results\cite{zhouPRB11} prior to publication.

We acknowledge financial support through Grant No. MAT2011-24331 (MEC), from JAE (CSIC) (M.B.),
and from ITN under Grant No. 234970 (EU). M. B. and A. P. J. are grateful to the FiDiPro program of the Academy
of Finland for support during the early stages of this project.
\end{acknowledgments}


\begin{widetext}
\appendix

\section{Derivation of the polarizability for circularly polarized field}
\label{appendix}

The derivation of the formula for the dynamical polarizability follows the lines of its counterpart in the 2DEG.\cite{chinodynscreeningPRB02} The wavefunction for graphene under a periodic driving can be written by use of the Floquet theorem as
\begin{align}
\psi_{\mathbf{k},\sigma}(\mathbf{r},t) = \frac{1}{\sqrt{2}}e^{i\mathbf{kr}}e^{-i\epsilon_{\mathbf{k},\sigma}t}\phi_{\mathbf{k},\sigma}(t),
\end{align}
where $\epsilon_{\mathbf{k},\sigma}$ is the quasienergy and $\phi_{\mathbf{k},\sigma}(t)$ are the Floquet states which fulfill the time-periodicity of the driving field, and we have chosen the solution corresponding to the First Brillouin zone. After applying a weak probe potential, these wavefunctions are not any more eigenfunctions of the full Hamiltonian, but we can use them as a basis to write the new wavefunction:
\begin{align}
\Psi_{\mathbf{k},\sigma}(\mathbf{r},t) = \sum_{\mathbf{k'}\sigma'}a_{\mathbf{k'},\sigma'}(t)\psi_{\mathbf{k'},\sigma'}(\mathbf{r},t)
\end{align}
Inserting this into the Schr\"{o}dinger equation for the Hamiltonian $H_0(t) + H_1(t)$, where $H_0(t)$ is the Hamiltonian for the periodically driven graphene, and $H_1(t) = V(\mathbf{r},t)$ represents the weak probe potential, we are left with a differential equation for the coefficients $a_{\mathbf{k},\sigma}(t)$:
\begin{align}
i\sum_{\mathbf{k'}\sigma'}\dot{a}_{\mathbf{k'},\sigma'}(t)\psi_{\mathbf{k'},\sigma'}(\mathbf{r},t) = \sum_{\mathbf{k'}\sigma'}a_{\mathbf{k'},\sigma'}(t)V(\mathbf{r},t)\psi_{\mathbf{k'},\sigma'}(\mathbf{r},t)
\end{align}
We can now project this equation into a state $\psi_{\mathbf{k''},\sigma''}$, yielding
\begin{equation}
\dot{a}_{\mathbf{k''},\sigma''}(t)
= -i\sum_{\mathbf{k'}\sigma'}a_{\mathbf{k'},\sigma'}(t)e^{i(\epsilon_{\mathbf{k''},\sigma''}-\epsilon_{\mathbf{k'},\sigma'})t}\phi_{\mathbf{k''},\sigma''}^{*}(t)\phi_{\mathbf{k'},\sigma'}(t)V(\mathbf{k''-k'},t),
\end{equation}
where $V(\mathbf{k''-k'},t)$ is the projection of the probe potential into the states $\mathbf{k'}$ and $\mathbf{k''}$.
Now we can expand this equation in a power series of the external potential, and keeping only the first order we are left with
\begin{align}
\dot{a}_{\mathbf{k''},\sigma''}^{(1)}(t) &= -ie^{i(\epsilon_{\mathbf{k''},\sigma''}-\epsilon_{\mathbf{k},\sigma})t}\phi_{\mathbf{k''},\sigma''}^{*}(t)\phi_{\mathbf{k},\sigma}(t)V(\mathbf{k''-k},t).
\end{align}
This equation can be simplified by Fourier transforming it, yielding
\begin{equation}
a_{\mathbf{k''},\sigma''}(t) = \int \frac{d\omega}{2\pi}V(\mathbf{k''-k},\omega)
e^{-i\omega t}e^{i(\epsilon_{\mathbf{k''},\sigma''}-\epsilon_{\mathbf{k},\sigma})t}e^{\eta t}\\
\sum_{nn'}\frac{e^{i(n'-n)\Omega t}\left[\phi_{\mathbf{k''},\sigma''}^{n',a*}\phi_{\mathbf{k},\sigma}^{n,a*}+\phi_{\mathbf{k''},\sigma''}^{n',b*}\phi_{\mathbf{k},\sigma}^{n,b*}\right]}{\omega-(n'-n)\Omega-(\epsilon_{\mathbf{k''},\sigma''}-\epsilon_{\mathbf{k},\sigma})+i\eta}.
\end{equation}

In order to get the response of the system to the external probe in linear response, we write down the expression of the induced charge density:
\begin{align}
\rho_{\mathbf{k},\sigma}^{\text{ind}}(\mathbf{r},t) &= \Psi_{\mathbf{k},\sigma}^{*}(\mathbf{r},t)\Psi_{\mathbf{k},\sigma}(\mathbf{r},t) - \psi_{\mathbf{k},\sigma}^{*}(\mathbf{r},t)\psi_{\mathbf{k},\sigma}(\mathbf{r},t)\nonumber\\
&= \sum_{\mathbf{k'}\sigma'}a_{\mathbf{k'},\sigma'}^{*}(t)\psi_{\mathbf{k'},\sigma'}^{*}(\mathbf{r},t)\psi_{\mathbf{k},\sigma}(\mathbf{r},t)+a_{\mathbf{k'},\sigma'}(t)\psi_{\mathbf{k},\sigma}^{*}(\mathbf{r},t)\psi_{\mathbf{k'},\sigma'}(\mathbf{r},t)
\end{align}
and insert the result obtained for $a_{\mathbf{k},\sigma}(t)$. After some algebra we arrive at
\begin{align}
\rho^{\text{ind}}(\mathbf{r},t) = \sum_{\mathbf{q}}\int\frac{d\omega}{2\pi}V^{\text{ext}}(\mathbf{q},\omega)e^{-i\omega t}e^{i\mathbf{qr}}\sum_{\sigma\sigma'}\sum_{\mathbf{k}}f_{\mathbf{k},\sigma}\mathcal{F}_{\mathbf{k},\sigma,\sigma'},
\end{align}
where we have introduced the short notation
\begin{align}
\mathcal{F}_{\mathbf{k},\sigma,\sigma'} = \sum_{nn'}\sum_{mm'}&\left[\frac{1}{2}\frac{e^{i(n'-n)\Omega t}e^{i(m'-m)\Omega t}
\left(\phi_{\mathbf{k+q},\sigma'}^{n',a*}\phi_{\mathbf{k},\sigma}^{n,a}+\phi_{\mathbf{k+q},\sigma'}^{n',b*}\phi_{\mathbf{k},\sigma}^{n,b}\right)
\left(\phi_{\mathbf{k},\sigma}^{m',a*}\phi_{\mathbf{k+q},\sigma'}^{m,a}+\phi_{\mathbf{k},\sigma}^{m',b*}\phi_{\mathbf{k+q},\sigma'}^{m,b}\right)}{\omega-(n'-n)\Omega-(\epsilon_{\mathbf{k+q},\sigma'}-\epsilon_{\mathbf{k},\sigma})+i\eta}\right.+ \nonumber\\
&\left.\frac{1}{2}\frac{e^{-i(n'-n)\Omega t}e^{-i(m'-m)\Omega t}
\left(\phi_{\mathbf{k-q},\sigma'}^{n',a}\phi_{\mathbf{k},\sigma}^{n,a*}+\phi_{\mathbf{k-q},\sigma'}^{n',b}\phi_{\mathbf{k},\sigma}^{n,b*}\right)
\left(\phi_{\mathbf{k},\sigma}^{m',a}\phi_{\mathbf{k-q},\sigma'}^{m, a*}+\phi_{\mathbf{k},\sigma}^{m',b}\phi_{\mathbf{k-q},\sigma'}^{m,b*}\right)}
{-\omega-(n'-n)\Omega-(\epsilon_{\mathbf{k-q},\sigma'}-\epsilon_{\mathbf{k},\sigma})-i\eta}\right].
\end{align}

By comparing with the Poisson equation
\begin{equation}
\rho^{\text{ind}}(\mathbf{r},t) = \sum_{\mathbf{q}}\int\frac{d\omega}{2\pi}V^{\text{ind}}(\mathbf{q},\omega)e^{-i\omega t}e^{i\mathbf{qr}}\frac{q^2}{4\pi}
\end{equation}
we see that the induced potential must fulfill
\begin{equation}
V^{\text{ind}}(\mathbf{q},\omega) = \frac{4\pi}{q^2}V^{\text{ext}}(\mathbf{q},\omega)\sum_{\sigma\sigma'}\sum_{\mathbf{k}}f_{\mathbf{k},\sigma}\mathcal{F}_{\mathbf{k},\sigma,\sigma'}.
\end{equation}
We sum on both sides $V^{\text{ext}}(\mathbf{q},\omega)$ and get
\begin{equation}
V^{\text{tot}}(\mathbf{q},\omega) = \left(1+\frac{4\pi}{q^2}\sum_{\sigma\sigma'}\sum_{\mathbf{k}}f_{\mathbf{k},\sigma}\mathcal{F}_{\mathbf{k},\sigma,\sigma'}\right)V^{\text{ext}}(\mathbf{q},\omega),
\end{equation}
where the dielectric function is given by
\begin{equation}
\varepsilon(\mathbf{q},\omega) = \frac{1}{1+\frac{4\pi}{q^2}\sum_{\sigma\sigma'}\sum_{\mathbf{k}}f_{\mathbf{k},\sigma}\mathcal{F}_{\mathbf{k},\sigma,\sigma'}}.
\end{equation}
In the RPA approximation we obtain therefore
\begin{equation}
\varepsilon(\mathbf{q},\omega)_{\text{RPA}} = 1 - \frac{4\pi}{q^2}
\sum_{\sigma\sigma'}\sum_{\mathbf{k}}f_{\mathbf{k},\sigma}\mathcal{F}_{\mathbf{k},\sigma,\sigma'}.
\end{equation}
After substituting the expression for $\mathcal{F}_{\mathbf{k},\sigma,\sigma'}$ and some straightforward manipulations, we arrive at our desired result for the dynamical polarizability:
\begin{equation}
\Pi(\mathbf{q},\omega) =  \sum_{\sigma\sigma'}\sum_{\mathbf{k}}\sum_{l}\frac{f_{\mathbf{k},\sigma}-f_{\mathbf{k+q},\sigma'}}{\omega  -\epsilon_{\mathbf{k+q},\sigma', l}+\epsilon_{\mathbf{k},\sigma,0}+i\eta}
\sum_{n} |\phi_{\mathbf{k+q},\sigma',l}^{n,a,*}\phi_{\mathbf{k},\sigma,0}^{n,a}+\phi_{\mathbf{k+q},\sigma',l}^{n,b,*}\phi_{\mathbf{k},\sigma,0}^{n,b}|^2
\end{equation}
Notice that now we have simplified the expression by writing it as the scalar product between different Floquet sidebands by using Eqs.~\eqref{transformation1} and \eqref{transformation2}.
\end{widetext}

\clearpage
\bibliography{bibliographie_graphene}{}

\end{document}